\documentclass[pra,aps,a4paper,showpacs,superscriptaddress,twocolumn]{revtex4}
\usepackage{amsmath}
\usepackage{amsfonts}
\usepackage{amssymb}
\usepackage{multirow}
\usepackage{graphicx}
\usepackage{dcolumn}
\begin{document}

\title{Ground state of spin-1 Bose-Einstein condensates with spin-orbit coupling in a Zeeman field}
\author{L. Wen}
\affiliation {Institute of Physics, Chinese Academy of
Sciences, Beijing 100080, China}
\author{Q. Sun}\email{sunqing@iphy.ac.cn}
\affiliation {Institute of Physics, Chinese Academy of
Sciences, Beijing 100080, China}
\author{H. Q. Wang}
\affiliation {School of Statistics and Mathematics, Yunnan University of
Finance and Economics, Kunming, Yunnan Province, 650221, China}
\author{A. C. Ji}\email{andrewjee@sina.com}
\affiliation {Center of Theoretical Physics, Department of Physics, Capital Normal University, Beijing 100048, China}
\author{W.~M. Liu}
\affiliation {Institute of Physics, Chinese Academy of
Sciences, Beijing 100080, China}
\begin{abstract}
We systematically investigate the weakly trapped spin-1 Bose-Einstein condensates with spin-orbit coupling in an external Zeeman field. We find that the mean-field ground state favors either a magnetized standing wave phase or plane wave phase when the strength of Zeeman field is below a critical value related to the strength of spin-orbit coupling. Zeeman field can induce the phase transition between standing wave and plane wave phases, and we determine the phase boundary analytically and numerically. The magnetization of these two phases responds to the external magnetic field in a very unique manner, the linear Zeeman effect magnetizes the standing wave phase along the direction of the magnetic field, but the quadratic one demagnetizes the plane wave phase. When the strength of Zeeman field surpasses the critical value, the system is completely polarized to a ferromagnetic state or polar state with zero momentum.
\end{abstract}

\pacs{05.30.Jp, 03.75.Mn, 67.85.Fg, 67.85.Jk}
\maketitle

\section{Introduction}
Spin-orbit coupling, which is generally referred to the coupling of a particle's spin with its degree of motion in quantum physics, is responsible for many fundamental physical phenomena in quantum systems, such as the spin Hall effects and topological insulators \cite{Top}. Recently, with the pioneering experimental realization of an artificial abelian or non-abelian gauge potential in neutral atoms \cite{gauge potential1,gauge potential2}, the effective spin-orbit coupling has been created in spinor Bose-Einstein condensates (BEC) by dressing two atomic spin states with a pair of lasers \cite{experiment1,experiment2,experiment3,experiment4}, and has attracted a great deal of attention in the condensed matter community. Especially, the spin-orbit coupling effects can give rise to many intriguing exotic ground states in an interacting ultracold spinor Bose gas, such as the plane wave (PW), standing wave (SW), triangular lattice and square lattice phases \cite{T1,ZhaiHui,T2,T3,T4,T5,T6,T7,T8,T9,T10,T11,T12,T13,T14,T15,T16,T17,T18,T19,T20}. Moreover, in pseudo spin-1/2 BEC, the combined effects of Rashba spin-orbit coupling and rotating trap can also produce some unusual topological patterns including the skyrmion and giant vortex \cite{skyrmion1,skyrmion2,skyrmion3}.

Generally, without spin-orbit coupling, a spinor BEC exhibits a variety of magnetic phenomena. Taking the spin-1 BEC as an example, the mean-field ground state can be a ferromagnetic, antiferromagnetic or polar states \cite{TH}. Which type of these phases is favored by the ground state depends on the spin-dependent atomic interaction. Interestingly, an external Zeeman field can transform these phases into each other via tuning their magnetization, and results in a rich ground state phase diagram \cite{DW,Ueda}. For example, a broken-axisymmetry state can emerge as an intermediate phase in the transition between the ferromagnetic and polar phases. Such phase transitions reflect that the magnetic system responds to the external magnetic field in a very unique manner. So the application of an external magnetic field provides a powerful way to experimentally manipulate the magnetic behaviors of a spinor BEC \cite{DW}.

However, when the spin-orbit coupling is taken into account, the situation changes. The ground state of spin-1 BEC now could be a PW phase occupying a single momentum state or a SW phase formed by the coherent superposition of two plane waves with opposite momentum\cite{ZhaiHui}, and these two phases possess different magnetic properties. So it is interesting and important to study the effects of an external Zeeman field on these phases, which is the main context of this work. To make this point more transparent, in this paper, we consider a weakly trapped two-dimensional spin-1 BEC with Rashba spin-orbit coupling in a Zeeman field. We find that, if the Zeeman field is not too strong, the system is in a magnetized SW phase or PW phase, and the phase boundary immerses into SW phase due to the competition between Zeeman effects and spin-dependent atomic interaction. The Zeeman field modulates the magnetization of these two phases in different manners. For example, the linear Zeeman field magnetizes the SW phase, but the quadratic one demagnetizes the PW phase. While the strength of the Zeeman field surpasses a critical value determined by the strength of spin-orbit coupling, the system is completely polarized to either a ferromagnetic state or a polar state with zero momentum.

This paper is organized as follows. We first give the model in Sec. $\textrm{II}$, and discuss the single-particle ground state in Sec. $\textrm{III}$. By using the variational approximation method and the numerical simulation, we give the phase diagram of mean-field ground state and study the effects of Zeeman field on the magnetization of this system in Sec. $\textrm{IV}$. Finally, we conclude our results in Sec. $\textrm{V}$.

\section{The Model}
We consider a quasi-two-dimensional spin-1 BEC with Rashba type spin-orbit coupling in the Zeeman field, where the particle's spin couples its degree of motion in $xy$ plane. In the mean-field approximation, the Gross-Pitaevskii energy functional of such system is of the form
\begin{eqnarray}\label{E}
E&=&\int dr \bigg \{ \sum_{m}\psi_{m}\left[\!-\frac{\hbar^2\nabla^2}{2M}-pm+qm^2\right]\psi_{m} \notag  \\
\!\!&+&\!\!\!\frac{\hbar \kappa}{M}[\psi_{1}^{*}(-i \partial_x\!-\!\partial_y)\psi_0\!+\!\psi_0^{*} (-i \partial_x\!-\!\partial_y)\psi_{-1}\!+\!h.c.] \notag \\
\!&+&\!\!\!\frac{c_0}{2}n^2\!+\!\frac{c_2}{2}[(n_{1}-n_{-1})^2\!+\!2|\psi_{1}^{*}\psi_0+\psi_0^{*}\psi_{-1}|^2]\bigg\},
\end{eqnarray}
where $m=1,0,-1$ and the density distribution of the component $m$ reads $n_m=\left\vert\psi_m\right\vert^2$ with the condensation wave function $\psi_m$, thus the total atomic density is $n=\sum_{m} n_{m}$. For the spin-orbit coupling term, we consider the symmetric Rashba case $\kappa_x=\kappa_y=\kappa$, where $\kappa$ is the strength of spin-orbit coupling. For the interaction terms, $c_0$ and $c_2$ are the interaction parameters which depend on two-body s-wave scattering lengths $a_0$ and $a_2$ for total spin 0,2: $c_0=4\pi\hbar^2 N(a_0+2a_2)/3M$ and $c_2=4\pi\hbar^2(a_2-a_0)/3M$ with the atomic mass $M$ and total atomic number $N$. To avoid the collapse of BEC under attractive interaction, $c_0$ is assumed to be non-negative throughout this paper. Moreover, $p$ and $q$ represent the strength of linear and quadratic Zeeman effects, respectively. For the sake of simplicity, the external magnetic field is assumed to be applied in the z-direction, and the values of $p$ and $q$ are taken to be non-negative constants.

\section{single-particle ground state}
\begin{figure}[tbp]
\includegraphics[ width= 0.48\textwidth]{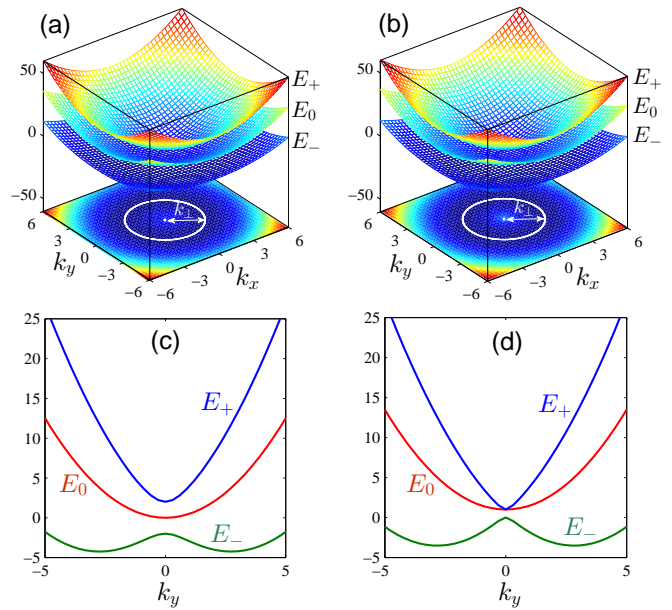}
\caption{(Color online) (a) The single-particle energy spectrum in linear Zeeman field with $p=2$. (b) The single-particle energy spectrum in quadratic Zeeman field with $q=1$. Noting that the figures in the bottom of (a) and (b) represent the projections of $E_-$ on the $k_x$-$k_y$ plane, and the white circular rings are the momentum of infinitely degenerate ground states in $E_-$. Furthermore, to clarify the energy gap between these energy branches, figures (c) and (d) have plotted the energy spectrums at $k_x=0$. In all the figures, the other parameters are the same that $M=\hbar=1$ and $\kappa=2$.  \label{fig1}}
\end{figure}
To gain some intuitions, it is instructive to start our investigations with the noninteracting Hamiltonian in the homogeneous case. The single-particle Hamiltonian preserves the symmetries of simultaneous spin and space rotations around the direction of applied magnetic field. The group describing such symmetries is $SO(2)$ which has discrete subgroups $\mathcal{C}_{nz}$ \cite{T17}. The single-particle ground state in absence of the external magnetic field is not unique so that it is infinitely degenerate along a circular ring with radius $k_\bot=\sqrt{k_x^2+k_y^2}= \sqrt{2}\kappa/\hbar$ in two-dimensional momentum space \cite{ZhaiHui,T18}, where $k_x$ and $k_y$ are the components of momentum along $x$ and $y$ directions, respectively. In presence of the Zeeman field, we first consider two simple cases in which the linear and quadratic Zeeman effects are considered separately. The combinative effects of them will be considered later.

Due to the typical Rashba spin-orbit coupling, the single-particle energy spectrums in momentum space can be obtained by virtue of the exact diagonalization scheme as follows
\begin{subequations}\label{ES}
\begin{eqnarray}
\!\!\!\!\!\!\!\!L:E_{0}\!\!&=&\!\!E_{\mathbf{k}} \text{,} \ E_{\pm }\!\!=\!\!E_{\mathbf{k}}\!\pm\! \sqrt{p^{2}+\frac{4\kappa^2}{M}E_{\mathbf{k}}}, \\
\!\!\!\!\!\!\!\!Q:E_{0}\!\!&=&\!\!E_{\mathbf{k}}\!+\!q \text{,}\ E_{\pm }\!\! =\!\!E_{\mathbf{k}}+\frac{1}{2}q\pm \frac{1}{2}\sqrt{q^{2}\!+\!\frac{16\kappa^2}{M}E_{\mathbf{k}} },
\end{eqnarray}
\end{subequations}
where $L$ and $Q$ represent the energy spectrums in linear and quadratic Zeeman fields, respectively. $E_{\mathbf{k}}=\hbar^2 k_\bot^2/2M$ is the kinetic energy of a free particle. The energy spectrums in Eq. (\ref{ES}) have been shown in Fig. \ref{fig1}(a) and \ref{fig1}(b), we can observe that the energy branch $E_{-}$ has lower energy no matter in the linear or quadratic Zeeman field. By minimizing the energy branch $E_-$ with respect to $k_\bot$, we can obtain the momentum of single-particle ground state which are $k_{\bot}=\sqrt{\frac{2\kappa ^{2}}{\hbar ^{2}}-\frac{M^{2}p^{2}}{2\hbar ^{2}\kappa ^{2}}}$ and $k_{\bot}= \sqrt{32\kappa^4-2q^2M^2}/(4\hbar \kappa)$, respectively. The corresponding single-particle eigenstates are given by
\begin{eqnarray}\label{solutions}
\Psi_{\mathbf{k}}\!\!=\!\!\left(\begin{array}{c}
         \alpha_1  \\
         -\alpha_0 e^{i\theta}\\
         \alpha_{-1} e^{i2\theta}
       \end{array}\right)e^{i\mathbf{k}\cdot{\mathbf{r}}},
\end{eqnarray}
where $\theta=\arctan{(k_y/k_x)}$, and the coefficients $\alpha_m$ satisfying $\sum_m \alpha_m^2=1$ are
\begin{subequations}\label{coefficients1}
\begin{eqnarray}
\!\!\!\!\!\!L&:&\alpha _{1}\!\!=\!\!\frac{2\kappa^2\!+\!Mp}{4\kappa^2},\alpha _{-1}\!\!=\!\!\frac{2\kappa ^{2}\!-\!M p}{4\kappa ^{2}},\alpha _{0}\!\!=\!\!\sqrt{2\alpha_1\alpha_{-1}},  \\
\!\!\!\!\!\!Q&:&\alpha _{1}\!\!=\!\!\alpha_{-1}\!\!=\!\!\frac{\sqrt{4\kappa^2-M q}}{4\kappa},\alpha _{0}\!\!=\!\! \frac{\sqrt{2}\sqrt{4\kappa^2+M q}}{4\kappa}.
\end{eqnarray}
\end{subequations}

From the results above, on the one hand, the Zeeman effects shift the momentum of ground state compared to the case in absence of external magnetic field. All the single-particle states with same momentum $k_{\bot}$ but different azimuthal angle in two-dimensional momentum space are degenerate ground states, namely, the single-particle ground states are infinitely degenerate along a circular ring with the radius $k_\bot$ shown in Fig. \ref{fig1}(a) and \ref{fig1}(b). Moreover, since the momentum of ground state should be real, we have noted that there exists two additional restraint conditions $p\leq 2\kappa^2/M$ and $q\leq 4\kappa^2/M$. In particular, if the strength of linear or quadratic Zeeman effect takes the maximal critical value, the circular ring of momentum in infinitely degenerate ground states shrinks to the zero momentum point (the center of ring shown in Fig. \ref{fig1}(a) and \ref{fig1}(b)) so that the system is polarized to a non-degenerate ground state. On the other hand, the presence of Zeeman field can open a energy gap between these energy branches as shown in Fig. \ref{fig1}(c) and \ref{fig1}(d). In linear Zeeman field, the gap between adjacent energy branches are $\Delta_{+0}=E_+-E_{0}=p$ and $\Delta_{0-}=E_0-E_{-}=\kappa^2/M+Mp^2/4\kappa^2$. While the quadratic Zeeman field only opens the gap $\Delta_{+-}=\Delta_{0-}=q/2+\kappa^2/M+Mq^2/16\kappa^2$ between the energy branches $E_+$ (or $E_0$) and $E_{-}$, it is gapless between the energy branches $E_+$ and $E_0$. These facts demonstrate that the linear and quadratic Zeeman effects play different roles on affecting the structure of single-particle energy spectrum.
\begin{figure}
\includegraphics[width= 0.48\textwidth]{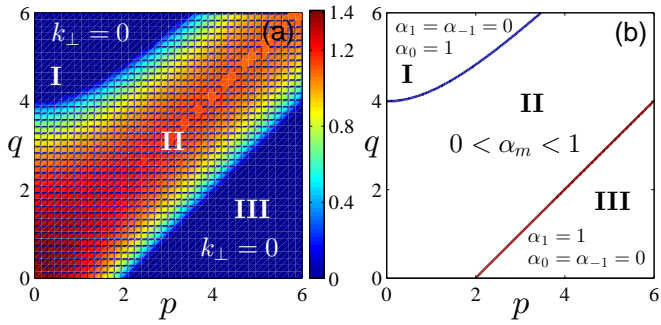}%
\caption{(Color online) (a) The momentum of single-particle ground state changes with linear and quadratic Zeeman effects. (b) The coefficients $\alpha_m$ of single-particle eigenstates in Eq. (\ref{solutions}) change with the linear and quadratic Zeeman effects. In both (a) and (b), the parameters are $\kappa=1$ and $\hbar=M=1$. \label{fig2}}
\end{figure}

Let us now consider the combination effect of the linear and quadratic Zeeman terms. For simplification, we just give the lower energy branch in single-particle energy spectrums as follows
\begin{equation}\label{spectrum_pq}
E_{-}=E_{\mathbf{k}}+\frac{2q}{3}+\omega\sqrt[3]{\sqrt{\Lambda}-\Theta}
+\omega^2\sqrt[3]{-\sqrt{\Lambda}-\Theta},
\end{equation}
where $\omega =-\exp{(i\pi/3)}$, $\Lambda =\Theta ^{2}-\Gamma ^{3}$ with $\Theta =q( q^{2}+18\kappa^2 E_{\mathbf{k}}/m-9p^{2})/27 $ and $\Gamma
=(3p^{2}+q^{2}+12\kappa ^{2}E_{\mathbf{k}}/m)/9$. We have to point out that $p$ and $q$ must satisfy $\Lambda \leq 0$ for the real energy eignvalue $E_{-}$. Since the expression of energy spectrum in Eq. (\ref{spectrum_pq}) is very complex, it is impossible to analytically solve the momentum of single-particle ground state and derive the coefficient $\alpha_m$ of the corresponding eigenstates in Eq. (\ref{solutions}), so we plot this quantities in Fig. \ref{fig2} by seeking the minimal value of energy Eq. (\ref{spectrum_pq}) numerically. It can be clearly seen that the quadratic Zeeman effect dominates in region $\textrm{I}$, but the linear Zeeman effect dominates in region $\textrm{III}$. In both the regions $\textrm{I}$ and $\textrm{III}$, the momentum of single-particle ground state is zero, the corresponding coefficients $\alpha_m$ in single-particle eigenstate are $\left(0,1,0\right)^T$ and $\left(1,0,0\right)^T$, respectively. These results imply that the system is polarized by the magnetic field in both the regions $\textrm{I}$ and $\textrm{III}$. While in region $\textrm{II}$, the coefficient satisfies $0<\alpha_m<1$, and the momentum of ground state is nonzero so that the single-particle ground states are infinitely degenerate along a circular ring with radius $k_\bot$ in momentum space. Furthermore, the boundary between the regions $\textrm{II}$ and $\textrm{III}$ can be analytically fixed that is $q/3-p-\zeta=0$ with $\zeta=\omega\sqrt[3]{\sqrt{\Lambda}-\Theta}
+\omega^2\sqrt[3]{-\sqrt{\Lambda}-\Theta}$.

\section{mean-field ground state in external Zeeman field}
For the spin-1 BEC with Rashba spin-orbit coupling, the interaction effects in absence of Zeeman field have been investigated extensively \cite{T1,ZhaiHui,T2,T3,T4,T5,T6,T7,T8,T9,T10,T11,T12,T13,T14,T15,T16,T17,T18,T19,T20}. In homogeneous case, the mean-field ground state favors either the PW for $c_2<0$ or SW for $c_2>0$, the former is ferromagnetic state, but the latter is polar state \cite{ZhaiHui}. These phases spontaneously break the rotational symmetry around the $z$ direction despite the fact that the Hamiltonian is axisymmetric. In particular, these two different phases are degenerate for $c_2=0$. Now the point is that if a Zeeman field along $z$-direction is turned on, we are inquisitive about whether an external magnetic field can essentially change the phase diagram of mean-field ground state for spin-orbit coupled spin-1 BEC? How the external magnetic field affects the magnetization of ground state? The investigation of the unique features of these problems is the primary purpose of this section.

In two dimensional homogeneous system, the interacting Hamiltonian preserves the symmetries of single-particle Hamiltonian, and the atomic interactions can couple different single-particle states so that the ground state preserving symmetry $\mathcal{C}_{nz}$ can be approximately described by the linear superposition of single-particle eigenstates on the degenerate momentum ring, $\Psi=\sum_n \mathcal{A}_n \Psi_{\mathbf{k}}$ with $\mathcal{A}_n$ satisfying $\sum_n \left\vert \mathcal{A}_n\right\vert^2=1$, where $n$ is an non-negative integer \cite{ZhaiHui,T17}. By substituting the ansatz into Eq. (\ref{E}) to compute the energy, we find that these high-symmetry states with $n\geq 3$ are energetically unfavored by ground state which can be verified by our numerical results. Thus this fact motivates us to choose the following ansatz for clarifying the phase diagram of mean-field ground state and developing a simple physical understanding in below discussions
\begin{equation}\label{ansatz1}
\Psi=\mathcal{A}\Psi_{\mathbf{k}}+\mathcal{B}\Psi_{-\mathbf{k}},
\end{equation}
where $\Psi_{\mathbf{k}}$ and $\Psi_{\mathbf{-k}}$ are two counter-propagating plane waves in Eq. (\ref{solutions}), and the superposition coefficients $\mathcal{A}$ and $\mathcal{B}$ are real constants satisfying the normalization condition $\mathcal{A}^2+\mathcal{B}^2=1$ under $\sum_{m}\int \left\vert \psi_m\right\vert^2 dr=1$.
\begin{figure}[tbp]
\includegraphics[width= 0.48\textwidth]{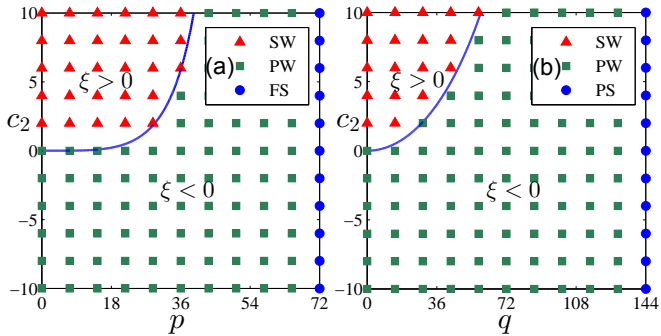}
\caption{(Color online) (a) The phase diagram of mean-field ground state in linear Zeeman field. (b) The phase diagram of mean-field in quadratic Zeeman field. The red triangles and green squares respectively denote the SW and PW phases given by numerical results, and the blue line is the phase boundary $\xi=0$ between these two phases. The blue circles represent the zero momentum ferromagnetic state (FS) in (a) and the zero momentum polar state (PS) in (b), respectively. The other parameters in both the figures are $c_0=100$, $M=1$ and $\kappa=6$. \label{fig3}}
\end{figure}

By inserting the ansatz into the mean-field energy functional Eq. (\ref{E}), we can obtain the energy expression as a function of the parameter $\mathcal{A}$
\begin{eqnarray}\label{E1}
E[\mathcal{A}]=\xi\left(\mathcal{A}^4-\mathcal{A}^2\right)+\text{const},
\end{eqnarray}
where $\xi=[2\alpha_0^2(\alpha_1^2+6\alpha_1\alpha_{-1}+\alpha_{-1}^2)-(\alpha_1^2-\alpha_{-1}^2)^2]c_2-(\alpha_1^2-\alpha_0^2+\alpha_{-1}^2)^2 c_0$.
Noting that some constants unrelated to $\mathcal{A}$ in Eq. (\ref{E1}) are omitted. We can immediately determine the phase diagram of mean-field ground state by minimizing the energy with respect to $\mathcal{A}$: (i) The PW phase with $\mathcal{A}=0$ and $\mathcal{B}=1$ if $\xi<0$; (ii) The SW phase with $\mathcal{A}=\mathcal{B}=1/\sqrt{2}$ if $\xi>0$. In particular, these two phases are degenerate when $\xi=0$. These results demonstrate that the external magnetic field shifts the phase boundary from $c_2=0$ to $\xi(c_2,p,q)=0$ due to the competition between the Zeeman effects and spin-dependent interaction.

To further explore the physics behind this, in what follows, we perform the detailed analysis on the phase diagram of ground state, and investigate the effects of external magnetic field on the magnetic behaviors of this system. We first consider the linear and quadratic Zeeman effects, respectively. The combination effects of them will be considered afterwards.

\subsection{Phase diagram and magnetic behaviors of ground state in linear or quadratic Zeeman field}
In the linear or quadratic Zeeman field, the phase boundary are $\xi=\left( 2-M^{2}p^{2}/\kappa ^{4}\right)c_{2} +\frac{M^{4}p^{4}}{16{\kappa }^{8}}\left(c_{2}-c_{0}\right)$ or $\xi=2c_2-\frac{M^2q^2}{16\kappa^4}\left(2c_2+c_0\right)$, respectively. In parameter spaces of ($c_2$,$p$) and ($c_2$,$q$), the corresponding phase diagrams of mean-field ground state have been shown in Fig. \ref{fig3}. We can observe that the SW phase only exists in the case of $c_2>0$, and the region of which is less than the region occupied by the PW in the $c_2$-$p$ or $c_2$-$q$ plane. More importantly, the external magnetic field can induce the phase transition from the SW to PW when the strength of magnetic field surpasses the critical value $p_0=2\kappa^2 \sqrt{(\sqrt{2c_2^2+2c_2c_0}-2c_2)/[M^2(c_0-c_2)]}$ on the phase boundary for the linear Zeeman effect (or $q_0=4\kappa^2\sqrt{2c_2/[(c_0+2c_2)M^2]}$ for the quadratic one), where we require $c_0\gg c_2$ in accordance with the real experiments. In particular, if the strength of linear (or quadratic) Zeeman effect attain the critical values $p_m=2\kappa^2/M$ (or $q_m=4\kappa^2/M$), the system can be polarized by the magnetic field so that all the atoms will occupy the state $\psi_1$ (or $\psi_0$) with zero momentum, in which the ground state is ferromagnetic state (or polar state).

Based on the coupled Gross-Pitaevskii equations, we have checked all the analytical predictions by using the imaginary time evolution
method to numerically look for the ground state solutions which can minimize the mean-field energy. We would like to point out that a very weak isotropic harmonic potential is included in our numerical simulation, it does not change the results in the homogeneous case. As shown in Fig. \ref{fig3}, these analytical predictions given by the ansatz in Eq. (\ref{ansatz1}) agree with the numerical results very well, which demonstrates that the excellent ansatz in Eq. (\ref{ansatz1}) captures all the fundamental physics.

In hindsight, the shifts of phase boundary can be understood by analyzing the magnetization of ground state in response to the external magnetic field. Generally, the magnetization for spin-1 BEC is defined as $\mathcal{M}=\sqrt{\mathcal{M}_x^2+\mathcal{M}_y^2+\mathcal{M}_z^2}$ with $\mathcal{M}_{\mu}=\sum_{mn}\int \psi_m^{*}(\sigma_{\mu})_{mn} \psi_n dxdy$ being the component of magnetization along the directions $\mu=x,y,z$, where the subscripts $m,n=1,0,-1$, and $\sigma_{\mu}$ represents the $3\times 3$ spin matrix for spin-1 \cite{Ueda}. A priori, we can use the ansatz in Eq. (\ref{ansatz1}) to analytically compute the magnitude of magnetization for the cases of $c_2<0$ and $c_2>0$, respectively. The results are listed in the table (\ref{Tab1}). It can be seen that the SW and PW phases respond to the external magnetic field in very different manners. On the one hand, in linear Zeeman field, the magnitude of magnetization of PW phase for $c_2<0$ always keeps invariant regardless of the external magnetic field. While in the regions $p<p_0$ and $c_2>0$, the system is initially in SW phase, the magnitude of magnetization increases linearly from zero until $p>p_0$, in which the system has transformed into PW and the magnitude of magnetization undergoes a discontinuous sudden transition from $\mathcal{M}=Mp_0/(2\kappa^2)$ to $\mathcal{M}=1$. On the other hand, an opposite story occurs in quadratic Zeeman field. With increasing the strength of magnetic field, the magnitude of magnetization of PW phase for $c_2<0$ decays quadratically from $\mathcal{M}=1$ to $\mathcal{M}=0$, which implies that the quadratic Zeeman effect demagnetizes the PW. However, in SW phase for $q<q_0$ and $c_2>0$, the quadratic Zeeman effect can not change its magnetization with $\mathcal{M}=0$. When the strength of magnetic field surpasses the critical value $q_0$, the system suddenly possesses magnetization $\mathcal{M}=\sqrt{16\kappa^4-M^2q_0^2}/(4\kappa^2)$, which demonstrates that the phase transition from SW to PW occurs. Again, we have verified these predictions numerically and the results are summarized in Fig. \ref{fig4}.
{\renewcommand\arraystretch{1.8}
\begin{table}
\begin{tabular}{|c|c|c|}
\hline
 & Linear Zeeman field & Quadratic Zeeman field \\
\hline
\multirow{2}{*}{$c_2>0$} & $\mathcal{M}_{SW}^{p<p_0}=\frac{Mp}{2\kappa^2}$ & $\mathcal{M}_{SW}^{q<q_0}=0$  \\
\cline{2-3}
& $\mathcal{M}_{PW}^{p>p_0}=1$  & $\mathcal{M}_{PW}^{q>q_0}=\frac{\sqrt{16\kappa^4-M^2q^2}}{4\kappa^2}$\\
\hline
$c_2<0$ & $\mathcal{M}_{PW}=1$ & $\mathcal{M}_{PW}=\frac{\sqrt{16\kappa^4-M^2q^2}}{4\kappa^2}$ \\
\hline
\end{tabular}
\caption{The expressions of magnetization in linear and quadratic Zeeman fields, respectively. For $c_2>0$, the ground state during $p<p_0$ or $q<q_0$ is the SW phase, but the PW phase for $p>p_0$ or $q>q_0$. For $c_2<0$, the ground state is always the PW for an arbitrary $p$ or $q$. \label{Tab1}}
\end{table}
\begin{figure}[tbp]
\includegraphics[ width= 0.48\textwidth]{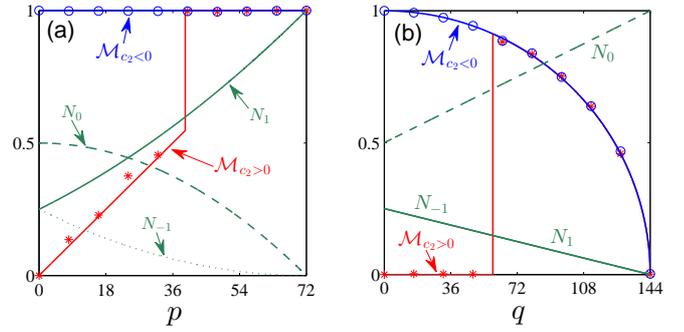}
\caption{(Color online) (a) The magnetization and atomic distribution of the ground state change with the strength of linear Zeeman field in the cases of $c_2=10$ and $c_2=-10$. (b) The magnetization and atomic distribution of the ground state change with the strength of quadratic Zeeman field in the cases of $c_2=10$ and $c_2=-10$. In both (a) and (b), the solid red and blue lines are respectively the corresponding magnetization of $c_2>0$ and $c_2<0$ given by the analytical results shown in table (\ref{Tab1}), the green lines represent the atomic numbers $N_m$. The red stars and blue circles given by our numerical results are the magnetization for $c_2>0$ and $c_2<0$, respectively. Moreover, the other parameters in both (a) and (b) are the same that $\kappa=6$, $\hbar=M=1$ and $c_0=100$. \label{fig4}}
\end{figure}

To highlight the effects of Zeeman field on the magnetization of ground state, further, we map the magnetization vectors of PW and SW phases onto a sphere with radius $\mathcal{M}=1$ as shown in Fig. \ref{fig5}. Before performing the analysis in detail, we introduce the polar angle $\vartheta = \arctan{\frac{\left\vert \mathcal{M}_+ \right\vert}{\mathcal{M}_z}}$ and azimuth angle $\varphi = \arctan{\frac{\mathcal{M}_y}{\mathcal{M}_z}}$ of the magnetization vector in spherical coordinate frame, where $\mathcal{M}_{+}=\mathcal{M}_{x}+i \mathcal{M}_y$. In presence of the external Zeeman field, the polar and azimuth angles of the magnetization vector for these two phases are
\begin{subequations}\label{angles}
\begin{eqnarray}
PW\!\!\!&:&\!\!\! \vartheta=\arctan{\frac{\sqrt{2}\alpha_0}{\alpha_1-\alpha_{-1}}},\quad \varphi=\theta,  \\
SW\!\!\!&:&\!\!\! \vartheta=\varphi=\text{undefined} \!\!\!\!\quad \text{for $p=0$ or $0\leq q<q_0$}, \\
     &&\!\!\! \vartheta=0,\varphi=\text{undefined} \!\!\!\!\quad \text{for $0<p<p_0$}, \notag \\
     && \!\!\!\vartheta=\frac{\pi}{2},\varphi=\theta \!\!\!\!\quad \text{for $q>q_0$},\notag
\end{eqnarray}
\end{subequations}
where $\theta=\arctan{(k_y/k_x)}$ is determined by the momentum of ground state. Motivated by the numerical results, we can fix $\theta=\frac{\pi}{4}$ in the following discussions. Noting that since the SW phase has no magnetization when $p=0$ or $0\leq q<q_0$, so the polar and azimuth angles of the magnetization vector for SW phase are undefined in the center of sphere. Furthermore, as we shall see below, the magnetization of SW phase is along the $z$-direction when $0<p<p_0$ in linear Zeeman field, so the azimuth angle of SW phase is undefined too. Fig. \ref{fig5} shows clearly that how the external magnetic field changes the magnetization of two phases. For the case of $c_2<0$, the ground state is PW, the linear Zeeman effect rotates the magnetization vector from transverse direction ($xy$ plane) to longitudinal direction ($z$ direction) along the meridian with azimuth angle $\varphi=\pi/4$, but the length of vector is invariant shown in Fig. \ref{fig5}(a). On the contrary, along the azimuth angle $\varphi=\pi/4$ direction, the quadratic Zeeman effect demagnetizes the PW in $xy$ plane by reducing the length of magnetization vector from one to zero as shown in Fig. \ref{fig5}(c). For the case of $c_2>0$ summarized in Fig. \ref{fig5}(b) and \ref{fig5}(d), the system would be initially in SW phase if $p<p_0$ or $q<q_0$. In this case, the length of magnetization vector in linear Zeeman field increases linearly along $z$ direction from zero, but it keeps invariant in quadratic one with $\mathcal{M}=0$. While $p$ (or $q$) surpasses the critical value $p_0$ (or $q_0$), the SW changes into the PW so that the magnetization vector responds to the Zeeman fields in the same manner as the one in PW.
\begin{figure}[tbp]
\includegraphics[ width= 0.45\textwidth]{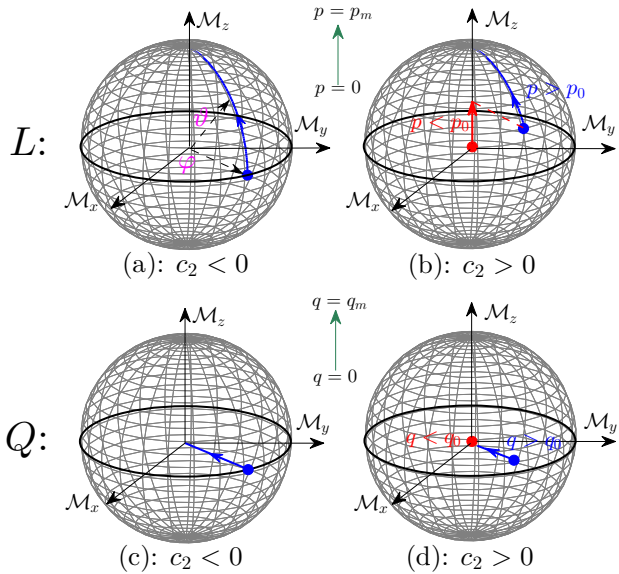}
\caption{(Color online) The magnetization vector responds to the linear (L) or quadratic (Q) Zeeman field. In (a)-(d), the red and blue solid circles represent the end of vectors for $c_2>0$ and $c_2<0$, respectively. Noting that the direction of the magnetization vector points to the red or blue circle from the center of sphere. Furthermore, the red and blue solid lines are the change trajectories of the end of magnetization vectors for SW and PW phases, respectively.  In linear Zeeman field, the end of vector for $c_2>0$ undergoes a transition from the $z$ direction to the blue line when $p>p_0$. While in quadratic Zeeman field, the end of the magnetization vector for $c_2>0$ undergoes the transition from the center of sphere to the azimuth angle direction when $q>q_0$. \label{fig5}}
\end{figure}
\subsection{Phase diagram and magnetic behaviors of ground state in the combination of linear and quadratic Zeeman fields}
So far, we have focused on the simple case that the linear and quadratic Zeeman effects are dealt with separately. Experimentally, the linear and quadratic Zeeman effects always coincide with the presence of external magnetic field. The former can be
effectively tuned by changing the total spin of the system, and the latter can be usually tuned by using a linearly polarized microwave field due to the alternating current Stark shift \cite{MP1,MP2}. As shown in Refs. \cite{DW,Ueda}, under a certain range of
linear and quadratic Zeeman effects, the ground state phase diagram of spin-1 BEC without spin-orbit coupling becomes much richer due to the competition between the linear and quadratic Zeeman effects and the spin-dependent interaction. The typical phase is the presence of a broken-axisymmetry state. Therefore, in this subsection we investigate the phase diagram of spin-1 BEC with spin-orbit coupling in the combination of linear and quadratic Zeeman effects.
\begin{figure}[tbp]
\includegraphics[ width= 0.48\textwidth]{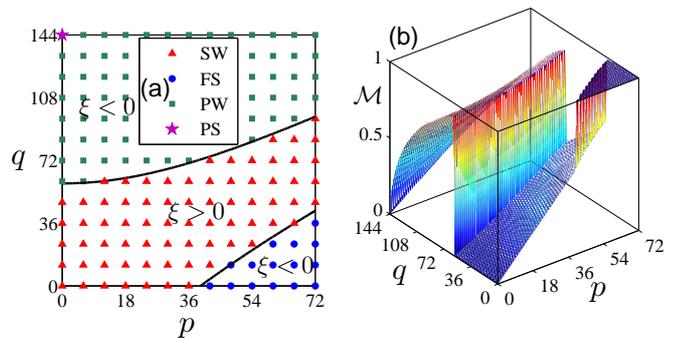}
\caption{(Color online) The phase diagram and magnetization distribution in the combination effects of linear and quadratic Zeeman fields. In the phase diagram (a), the red triangles, blue circles, green squares and purple stars respectively denote the SW, zero momentum ferromagnetic state (FS), PW and zero momentum polar state (PS) given by numerical results. The dark line is the phase boundary $\xi=0$ between these phases. (b) the magnetization distribution of the system versus $p$ and $q$. In both the figures, the parameters are $c_0=100$, $c_2=10$ and $\kappa=6$. \label{fig6}}
\end{figure}

We now have a similar situation as the single-particle problem in this combination effect. Since the expression of $\xi$ is very complex, it is impossible to analytically fix the phase boundary between the SW and PW phases. So we plot the phase diagram and magnetization distribution numerically in Fig. \ref{fig6} based on the ansatz Eq. (\ref{ansatz1}). It should be pointed out that for the case of $c_2<0$, the ground state is always the PW in the whole parameter space $\left(p,q\right)$ due to $\xi<0$. Thus we just consider the case of $c_2>0$ in following discussions for the sake of simplicity. From the phase diagram in Fig. \ref{fig6}(a), we can see that the ground state initially favors the SW phase when the strengths of linear and quadratic Zeeman effects are relative weak. However, the PW phase dominates if the quadratic Zeeman effect goes far beyond the linear one. Importantly, our numerical results show that all the atoms are polarized to the state $\psi_1$ for $q<14(p-39)/11$ with $p\in\left[39,p_m\right]$ or the state $\psi_0$ for $p=0$ and $q=q_m$, respectively. The corresponding order parameters of ground state in these regions are $\Psi_1=(1,0,0)^T$ and $\Psi_0=(0,1,0)^T$. In these states, the momentum of ground state is zero. All the results above can also be equally reflected by analyzing the magnetization of the system as shown in Fig. \ref{fig6}(b), which shows that the two states $\Psi_1$ and $\Psi_0$ are ferromagnetic state with $\mathcal{M}=1$ and polar state with $\mathcal{M}=0$, respectively.

\section{CONCLUSIONS}
Within the framework of the mean-field theory, we have systematically investigated the weakly trapped spin-1 Bose-Einstein condensates with spin-orbit coupling in an external Zeeman field, and clarified the effects of an external magnetic field on the system. In the single-particle case, the linear and quadratic Zeeman effects play the different roles on affecting the structure of single-particle energy spectrum. When the atomic interactions are taken into account, we give the phase diagram of mean-field ground state which is mainly comprised of the magnetized standing wave phase and plane wave phase. To develop a physical understanding, we use the linear superposition of two single-particle eigensates with counter-propagating wave vectors as the variational ansatz to analytically determine the phase boundary between these two phases, which agrees with our numerical results very well. The Zeeman field can induce the phase transition between the standing wave and plane wave phases, such phase transition can be further understood by analyzing the response of magnetization of these two phases to the external magnetic field. In particular, when the strength of Zeeman field surpasses a critical value which is related to the strength of spin-orbit coupling, the system is completely polarized to a ferromagnetic state or a polar state with zero momentum. These investigations not only help us to deepen the understanding of the physics behind the interaction between a matter field and gauge field, but also provide an effective way to manipulate the spin-orbit coupled spinor BEC experimentally by using an external magnetic field.

This work was supported by the NKBRSFC under Grants No. 2011CB921502, No. 2012CB821305, No. 2009CB930701, and No. 2010CB922904, NSFC under Grants No. 10934010, No. 11104064 and No. 60978019, No. 10901134, and NSFC-RGC under Grants No. 11061160490 and No. 1386-N-HKU748/10.

\end{document}